\documentclass[conference,a4paper,twocolumn]{IEEEtran}

\usepackage{lipsum}

\newcommand\blfootnote[1]{%
  \begingroup
  \renewcommand\thefootnote{}\footnote{#1}%
  \addtocounter{footnote}{-1}%
  \endgroup
}

\IEEEoverridecommandlockouts

\ifCLASSINFOpdf
\else
\fi
  \usepackage{tipa}
  \usepackage{graphicx}
	\usepackage{epsfig}
	\usepackage{epstopdf}
	\usepackage{upgreek}
	\usepackage[nolist,nohyperlinks]{acronym}

\usepackage[cmex10]{amsmath}

\usepackage{color, soul}

\usepackage{subfigure}

\usepackage{cite}

\begin{acronym}
\acro{NoC}{Network-on-Chip}
\acro{WNoC}{Wireless Network-on-Chip}
\acro{EM}{Electromagnetic}
\acro{SiO$_2$}{Silicon Dioxide}
\acro{AIN}{Aluminum nitride}
\acro{SiP}{System-in-Package}
\acro{SDM}{Software-defined metamaterial}
\end{acronym}

\begin{document}



\title{Modeling the EM Field Distribution within a Computer Chip Package}


\author{
\IEEEauthorblockN{
Xavier Timoneda, Sergi Abadal, Albert Cabellos-Aparicio, Eduard Alarc\'{o}n 
}
\IEEEauthorblockA{
\small{NaNoNetworking Center in Catalunya (N3Cat), Universitat Polit\`{e}cnica de Catalunya (UPC), Barcelona, Spain%
}}

\IEEEauthorblockA{
\IEEEauthorrefmark{1}\small{Email: xavier.timoneda@upc.edu
}
}
}


%


\maketitle


\begin{abstract}
Wireless Network-on-Chip (WNoC) appears as a promising alternative to conventional interconnect fabrics for chip-scale communications. The study of the channel inside the chip is essential to minimize latency and power. However, this requires long and computationally-intensive simulations which take a lot of time. We propose and implement an analytical model of the EM propagation inside the package based on ray tracing. This model could compute the electric field intensity inside the chip reducing the computational time several orders of magnitude with an average mismatch of only 1.7 dB.

\end{abstract}

\blfootnote{\textit{This work has been funded by the Spanish MINECO under grant PCIN-2015-012 and the EU under grant H2020-FETOPEN-736876.}} 



%
\IEEEpeerreviewmaketitle




\acresetall

\section{Introduction} \label{sec:introduction}
Network-on-Chip (NoC) has become the paradigm of choice to interconnect cores and memory within a Chip MultiProcessor (CMP). However, recent years have seen a significant increase in the number of cores per chip and, within this context, it becomes increasingly difficult to meet the communication requirements of CMPs with conventional NoCs alone. Advances in integrated mmWave antennas and transceivers have led to the proposal of Wireless Network-on-Chip (WNoC) as a potential alternative to conventional NoC fabrics. In a WNoC, certain cores are augmented with transceivers and antennas capable of modulating and radiating the information, which allow distant cores communicate with low latency and high throughput.

Due to its potential, WNoCs have been investigated extensively from the circuit \cite{Mondal2017}, and architecture perspectives \cite{AbadalASPLOS}. However, fewer works are focused on characterizing propagation within the computing package. Modeling the wireless channel is crucial to understand losses, dispersion, and multipath issues that impair communication and impact on the design and performance of the RF transceiver.

Manufacturing chips to experimentally evaluate the channel characteristics is very expensive in terms of time and money, so the majority of works studying the channel attenuation in a WNoC are performed using full-wave EM simulators over a virtual model of the package. However, these kind of simulations require large computational time specially for realistic and electrically large packages at high frequencies\cite{Timoneda2018}.

This article proposes an analytical model of the electromagnetic propagation inside the package based on ray tracing which takes into account the direct rays and the most significant multipath components. This model allows the user to obtain an accurate approximation of the field distribution over the package. The results are then compared with the ones obtained with a full-wave solver simulation of the same package.



\vspace{-0.1cm}
\section{System Model}
\label{sec:modeldescription}

\begin{figure}[!t]
\centering
\vspace{-0.2cm}
\includegraphics[width=0.85\columnwidth]{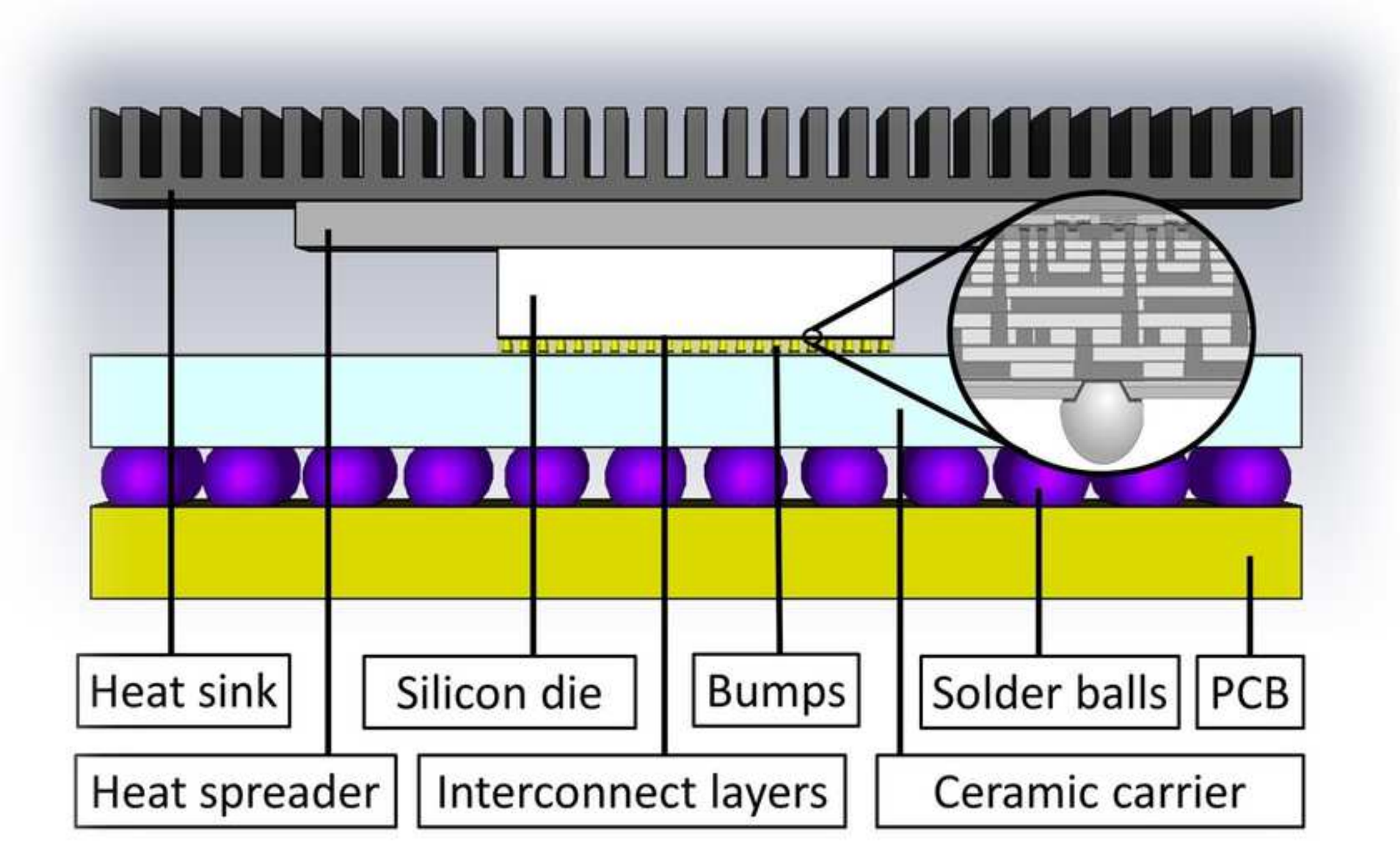}
\vspace{-0.2cm}
\caption{Schematic of the layers of a flip-chip package.}
\vspace{-0.4cm}
\label{fig:1}
\end{figure} 

The layers are described from top to bottom following Fig. \ref{fig:1}. On top, the heat sink and heat spreader dissipate the heat out of the silicon chip. Bulk silicon serves as the foundation of the transistors. The interconnect layers, are generally made of copper and surrounded by an insulator such as silicon dioxide (SiO$_2$) \cite{Markish2015}. The package carrier is a 0.5-mm thick ceramic, and at the bottom of it, an array of other solder balls is attached. 

\noindent
\textbf{Antenna:} The antenna used for the model is a square aperture antenna, which is modeled as an electrically small waveguide port. It is not a realistic antenna due to the impossibility of building such an ideal aperture, but it serves the purpose of channel characterization as it has a quasi-isotropic radiation diagram. Its radiation pattern was captured and used for the computation of the coefficients of the signal components in the model. The same procedure could be performed with other types of antenna.

\noindent
\textbf{Simulation model:} The structure shown in Figure \ref{fig:1} was introduced in a full-wave solver. The lateral dimensions for the silicon chip and the flip-chip ceramic package are 22 mm and 33 mm, respectively, which are typical values in both research and industry. The interconnect layer stack (13 $\upmu$m) is approximated by a SiO$_2$ layer under the silicon die and the antenna was placed on its center.

\vspace{-0.1cm}
\section{Analytical model}
\label{sec:analytical}
A ray tracing model was implemented, considering only the most significant narrow rays arriving to each point.
The multipath components taken into account are discussed below:

\noindent
\textbf{Direct Ray:} This is usually the strongest component and hence the most important for our model.

\noindent
\textbf{Reflected rays:} The only reflected rays taken into account are the ones reflected at the heatsink and the ones reflected once at the lateral boundaries of the chip. The rest do not have significant power to be considered. 

\noindent
\textbf{Diffracted rays:} The only diffracted rays taken into account are the ones which diffract on the Si layer to then reflect on the heatsink. The rest are blocked by the critical angle.

\noindent
\textbf{Scattering:} The scattering effect comes from rough surfaces. In our case, all the regions evaluated by the model have even surfaces. Thus, the scattering effect is not considered.

The electric field at an arbitrary point in the SiO$_2$ layer of the die can be expressed as the sum of the direct component plus the reflections at the heatsink and the edges of the chip. For modeling the change of amplitude depending of the distance of each ray, we will model the amplitude in a distance $d$ using the equation: $\frac{A_{0}}{A_{d}} = e^{\gamma d}$, where $\gamma$ is the propagation constant, and can be expressed as $\gamma = \alpha + j\beta$, where $\alpha$ is the attenuation constant, and $\beta$ is the phase constant.
The $E_{mag,d}$ at a given distance $d$, can be decomposed into real and imaginary parts as:
\begin{equation}
\label{eq:4}
E_{mag,d} = e^{-\alpha d} e^{-j\beta d}
\end{equation}
The constants $\alpha$ and $\beta$ can be expressed as:
\begin{equation}
\label{eq:5}
\alpha = \frac{\pi\sqrt{\epsilon_{r}}}{\lambda} tan(\delta),
\hspace{6mm}\beta = \frac{2\pi}{\lambda_{0}}\sqrt{\epsilon_{r}}
\end{equation}
Where tan⁡($\delta$) is the loss tangent of the material, and in the case of SiO$_2$ is 0.098 at 60 GHz and in the case of the Silicon is 0.252 at 60 GHz.

\subsection{Multipath components}
The electric field magnitude of the reflected rays at the heatsink can be expressed as a function of the direct ray as:
\begin{equation}
\label{eq:8}
E_{ref,heatsink} = T_{SiO2\_Si} R_{Al} T_{Si\_SiO2} E_{direct} 
\end{equation}
Where $T_{SiO2\_Si}$ is the transmission coefficient for each incidence angle between the SiO$_2$ and the Si layer, $R_{Al}$ is the reflection coefficient of the aluminum heatsink, and $T_{Si\_SiO2}$ is the transmission coefficient for each incidence angle between the Si and the SiO$_2$ layer. The same expression is used for computing the electric field magnitude of the reflected rays at the edges of the die.
For computing the transmission and reflection coefficients, equations derived from the Snell's law have been used, taking into account the incidence angle of each ray. These equations contemplate the possibility of having some rays whose transmitted angle is beyond the critical angle and therefore, their reflection coefficient is 1, keeping all the power in the same medium.

\subsection{Near field region}

It was observed that the near field region was the region with the biggest error using the attenuation formulas described above. This is because in the region between -1.3 and 1.3 mm, near field laws dominate the propagation ($\frac{\lambda_{0}}{\sqrt{\epsilon_{r}}} \approx 1.3 mm$). Therefore, specific equations according to \cite{Schantz2005} were used for this region in the model to reduce the error:
\begin{equation}
\label{eq:12}
P_{RX} \sim |E^{2}| \sim \frac{1}{(kd)^2} + \frac{1}{(kd)^4} + \frac{1}{(kd)^6}
\end{equation}
Where $k$ is the propagation constant and $d$ the distance from the source.

\vspace{-0.1cm}
\section{Results}
\label{sec:results}

\begin{figure}[!t]
\centering
\includegraphics[width=0.59\textwidth]{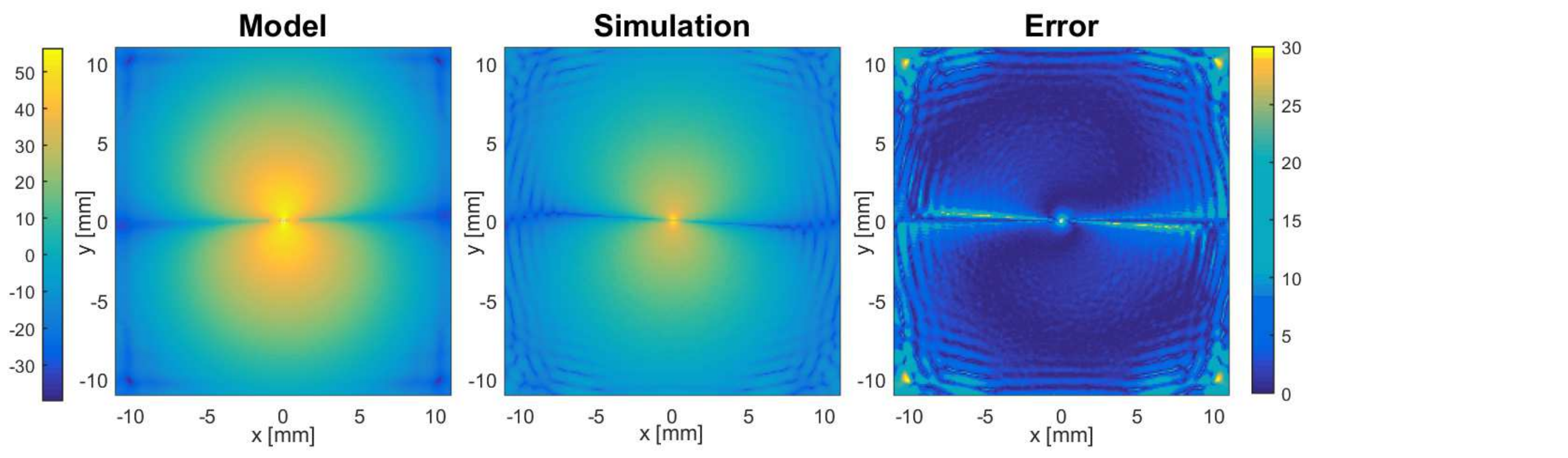}
\vspace{-0.2cm}
\caption{Top view of the electrical field computed using the model and the simulation results. Error between model and simulation.}
\vspace{-0.2cm}
\label{fig:2}
\end{figure}

Figure \ref{fig:2} shows the electric field magnitude computed using both analytical model and simulation. In the same figure, the error between both procedures was calculated. The edges of the die, and the region closer to the antenna have higher error, whereas in all the other regions the model matches very well, having a global geometric mean error of 1.7 dB.

\begin{figure}[!t]
\centering
\includegraphics[width=0.48\textwidth]{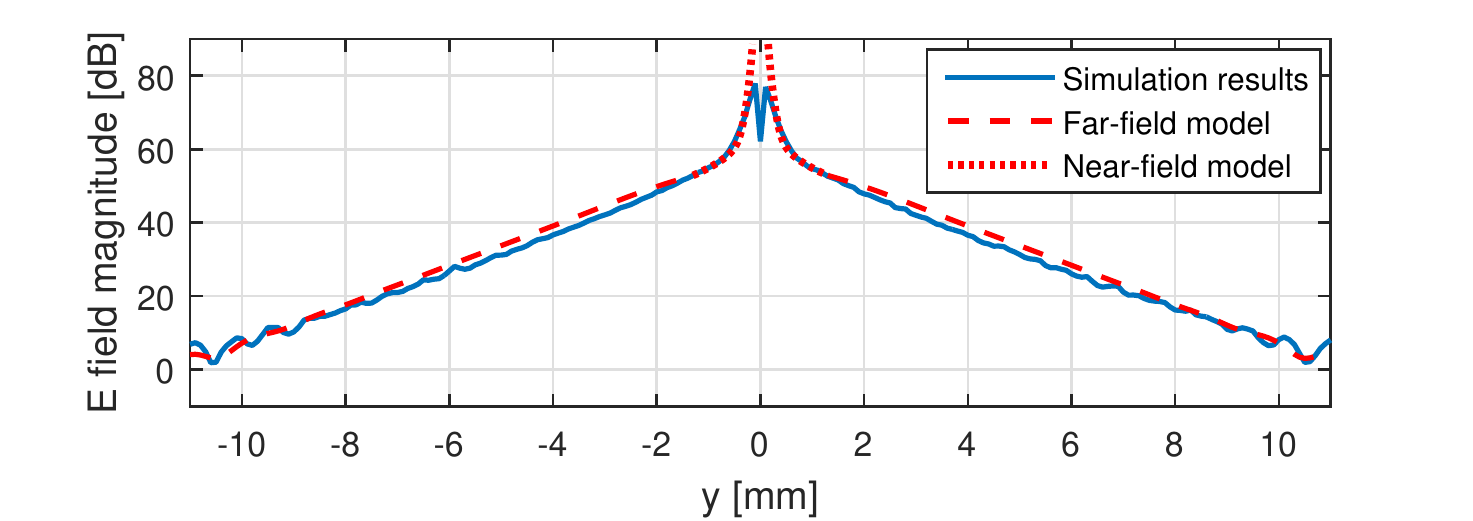}
\vspace{-0.2cm}
\caption{Magnitude of the electrical field in function of the distance with the antenna.}
\vspace{-0.2cm}
\label{fig:4}
\end{figure}

The error in the near field region was lowered by inserting the equations explained at Section 2 in the near field region. In Figure \ref{fig:4}  each of the fitting curves are superposed over the results and are only visible in their valid region.

\section{Conclusions}
\label{sec:conclusions}
In this work, we proposed a model based on ray tracing to study the EM propagation inside a computing package. Then, we created a script that implements an approximation using this model with high efficiency in terms of computation time and demonstrated its high accuracy.




\end{document}